\documentclass[aps,prc,twocolumn,superscriptaddress,floatfix]{revtex4-1}
\usepackage{textcomp}
\usepackage{nicefrac}
\usepackage{longtable}
\usepackage{graphicx}
\usepackage{color}            
\usepackage{multirow}
\usepackage{amsmath}
\usepackage{makecell}

\begin{document}
 
 \title{Charged-particle branching ratios above the neutron threshold in $^{19}$F: constraining $^{15}$N production in core-collapse supernovae.}
 
 \author{P. Adsley}
 \email{philip.adsley@wits.ac.za}
 \altaffiliation{Current Address: University of the Witwatersrand, South Africa and iThemba LABS, South Africa}
 \affiliation{Universit\'{e} Paris-Saclay, CNRS/IN2P3, IJCLab, 91405 Orsay, France}
 \author{F. Hammache}
 \email{hammache@ipno.in2p3.fr}
 \author{N. de S\'{e}r\'{e}ville}
 \affiliation{Universit\'{e} Paris-Saclay, CNRS/IN2P3, IJCLab, 91405 Orsay, France}
 \author{V. Alcindor}
 \affiliation{Universit\'{e} Paris-Saclay, CNRS/IN2P3, IJCLab, 91405 Orsay, France}
 \affiliation{GANIL, CEA/DRF-CNRS/IN2P3, Bvd Henri Becquerel, 14076 Caen, France}
 \author{M. Assi\'{e}}
 \author{D. Beaumel}
 \author{M. Chabot}
 \author{M. Degerlier}
 \author{C. Delafosse}
 \affiliation{Universit\'{e} Paris-Saclay, CNRS/IN2P3, IJCLab, 91405 Orsay, France}
 \author{T. Faestermann}
 \affiliation{Physik Department E12, Technische Universit\"{a}t M\"{u}nchen, D-85748 Garching, Germany}
 \author{F. Flavigny}
 \affiliation{Universit\'{e} Paris-Saclay, CNRS/IN2P3, IJCLab, 91405 Orsay, France}
 \author{S. P. Fox}
  \affiliation{Department of Physics, University of York, Heslington, York, YO10 5DD, United Kingdom}
 \author{R. Garg}
 \affiliation{Department of Physics, University of York, Heslington, York, YO10 5DD, United Kingdom}
 \affiliation{School of Physics and Astronomy, University of Edinburgh, Edinburgh, EH9 3FD, United Kingdom}
 \author{A. Georgiadou}
 \affiliation{Universit\'{e} Paris-Saclay, CNRS/IN2P3, IJCLab, 91405 Orsay, France}
 \author{S. A. Gillespie}
 \affiliation{Department of Physics, University of York, Heslington, York, YO10 5DD, United Kingdom}
  \altaffiliation{Current Address: TRIUMF, 4004 Wesbrook Mall, Vancouver, BC V6T 2A3, Canada}
 \author{J. Guillot}
  \affiliation{Universit\'{e} Paris-Saclay, CNRS/IN2P3, IJCLab, 91405 Orsay, France}
 \author{V. Guimar\~{a}es}
 \affiliation{Instituto de Fisica, Universidade de S\~{a}o Paulo, Rua do Mat\~{a}o, 1371, S\~{a}o Paulo 05508-090, SP, Brazil }
  \affiliation{Universit\'{e} Paris-Saclay, CNRS/IN2P3, IJCLab, 91405 Orsay, France}
 \author{A. Gottardo}
 \affiliation{Universit\'{e} Paris-Saclay, CNRS/IN2P3, IJCLab, 91405 Orsay, France}
 \author{R. Hertenberger}
 \affiliation{Fakult\"{a}t f\"{u}r Physik, Ludwig-Maximilians-Universit\"{a}t M\"{u}nchen, D-85748 Garching, Germany}
 \author{J. Kiener}
 \affiliation{Universit\'{e} Paris-Saclay, CNRS/IN2P3, IJCLab, 91405 Orsay, France}
 \author{A. M. Laird}
 \affiliation{Department of Physics, University of York, Heslington, York, YO10 5DD, United Kingdom}
 \author{A. Lefebvre-Schuhl}
 \affiliation{Universit\'{e} Paris-Saclay, CNRS/IN2P3, IJCLab, 91405 Orsay, France}
 \author{I. Matea}
 \author{A. Meyer}
 \affiliation{Universit\'{e} Paris-Saclay, CNRS/IN2P3, IJCLab, 91405 Orsay, France}
 \author{M. Mahgoub}
 \affiliation{Faculty of Science, Jazan University, Kingdom of Saudi Arabia}
 \affiliation{Sudan University of Science and Technology, Khartoum, Sudan}
 \author{L. Olivier}
 \author{L. Perrot}
 \affiliation{Universit\'{e} Paris-Saclay, CNRS/IN2P3, IJCLab, 91405 Orsay, France}
 \author{J. Riley}
 \affiliation{Department of Physics, University of York, Heslington, York, YO10 5DD, United Kingdom}
   \author{I. Sivacek}
 \affiliation{ASCR-Rez, CZ-250 68, Rez, Czech Republic}
 \author{I. Stefan}
 \affiliation{Universit\'{e} Paris-Saclay, CNRS/IN2P3, IJCLab, 91405 Orsay, France}
 \author{V. Tatischeff}
  \affiliation{Universit\'{e} Paris-Saclay, CNRS/IN2P3, IJCLab, 91405 Orsay, France}
  \author{H.-F. Wirth}
 \affiliation{Fakult\"{a}t f\"{u}r Physik, Ludwig-Maximilians-Universit\"{a}t M\"{u}nchen, D-85748 Garching, Germany}
  
 \date{\today}

\begin{abstract}

\begin{description}
\item[Background] Spatially-correlated overabundances of $^{15}$N and $^{18}$O observed in some low-density graphite meteoritic grains have been connected to nucleosynthesis taking place in the helium-burning shell during core-collapse supernovae. Two of the reactions which have been identified as important to the final abundances of $^{15}$N and $^{18}$O are $^{18}$F($n,\alpha$)$^{15}$N and $^{18}$F($n,p$)$^{18}$O.
\item[Purpose] The relative strengths of the $^{18}$F($n,\alpha$)$^{15}$N and $^{18}$F($n,p$)$^{18}$O reactions depend sensitively on the relative $\alpha_0$ and $p_0$ decay branches from states above the neutron threshold in $^{19}$F in addition to other properties such as the spins and parities, and the neutron widths. However, experimental data on the charged-particle decays from these highly excited states are lacking or inconsistent.
\item[Method] Two experiments were performed using proton inelastic scattering from LiF targets and magnetic spectrographs. The first experiment used the high-resolution Q3D spectrograph at Munich to constrain the properties of levels in $^{19}$F. A second experiment using the Orsay Split-Pole spectrograph and an array of silicon detectors was performed in order to measure the charged-particle decays of neutron-unbound levels in $^{19}$F.
\item[Results] A number of levels in $^{19}$F have been identified along with their corresponding charged-particle decays. The first state above the neutron threshold which has an observed proton-decay branch to the ground state of $^{18}$O lies 68 keV ($E_x = 10.5$ MeV) above the neutron threshold. The $\alpha$-particle decays from the neutron-unbound levels are generally observed to be much stronger than the proton decays.
\item[Conclusion] Neutron-unbound levels in $^{19}$F are observed to decay predominantly by $\alpha$-particle emission, supporting the role of $^{18}$F($n,\alpha$)$^{15}$N in the production of $^{15}$N in the helium-burning shell of supernovae. Improved resonant-scattering reaction data are required in order to be able to determine the reaction rates accurately.
\end{description}

\end{abstract}

\maketitle

\section{Astrophysical Background}

Recent analysis \cite{2041-8205-754-1-L8} of low-density graphite grains from the Orgueil meteorite show spatially-correlated excesses of $^{15}$N and $^{18}$O, suggesting a contribution of material originating from the inner part of the helium-rich zone of an exploding massive star \cite{PhysRevC.89.025807}. In this helium-rich layer, helium was being burnt into carbon and oxygen when the supernova occurred, causing a shockwave to pass through the outer layers of the star. The outer layers were heated and compressed, and were subsequently expelled into the interstellar medium, becoming incorporated in the graphite meteoritic grain.

At the end of hydrogen burning, some $^{14}$N is left as a result of the operation of the CNO cycles. This $^{14}$N is converted into $^{18}$F through the $^{14}$N($\alpha,\gamma$)$^{18}$F reaction. Under normal pre-supernova conditions, during which the mass fraction of neutrons is very low, this $^{18}$F undergoes $\beta^+$ decays into $^{18}$O. During the supernova explosion the $^{18}$O($\alpha,n$)$^{21}$Ne reaction begins to operate, supplying neutrons which can react with $^{18}$F, resulting in $^{18}$F($n,\alpha$)$^{15}$N and $^{18}$F($n,p$)$^{18}$O reactions. The temperatures in the helium-rich zone reach peak post-shock temperatures of $0.4-0.7$ GK, corresponding to a range of thermal energies of $kT = 35 - 60$ keV. The reaction rates depend on the properties of states within a few $kT$ of the neutron threshold, corresponding to approximately $E_{cm} < 200$ keV in the present case.

The recommendation of the sensitivity study of Bojazi and Meyer \cite{PhysRevC.89.025807} is that the focus of future evaluations of the $^{18}$F$+n$ reaction rates should attempt to describe the competition between the proton and $\alpha$-particle exit channels. This is particularly important because of the interplay between the two reactions: greater proton production from $^{18}$F($n,p$)$^{18}$O produces $^{15}$N via the $^{18}$F($n,p$)$^{18}$O($p,\alpha$)$^{15}$N reaction chain whilst also destroying it via proton absorption in
the $^{15}$N($p,\alpha$)$^{12}$C reaction. An increased production of $^{15}$N through the $^{18}$F($n,\alpha$)$^{15}$N reaction also results in a decreased destruction of $^{15}$N through the $^{15}$N($p,\alpha$)$^{12}$C reaction enabled by the protons produced in the competing $^{18}$F($n,p$)$^{18}$O reaction. After the supernova shockwave has passed the remaining $^{18}$F will decay into $^{18}$O.

Presently the $^{18}$F($n,\alpha$)$^{15}$N and $^{18}$F($n,p$)$^{18}$O reaction rates are based on Hauser-Feshbach calculations \cite{CAUGHLAN1988283,vangioni1986advances}. However, the level density at the neutron threshold in the compound nucleus $^{19}$F ($S_n = 10.432$ MeV) may not be high enough for statistical models to be used with a great deal of confidence. For example, the study of the $^{26}$Al($n,p$)$^{26}$Mg and $^{26}$Al($n,\alpha$)$^{23}$Na reactions by Koehler {\it et al.} \cite{PhysRevC.56.1138} found that there were significant disagreements between the measured reaction rates and those based on statistical models. The neutron energies which dominate the neutron-induced reactions are spread over a smaller energy range than for changed-particle reactions. Therefore, he number of levels which contribute to the reaction at each temperature is lower for neutron-induced reactions compared to charged particle-induced reactions.

Additional factors may also lead one to conclude that statistical models may be inappropriate for computing the $^{18}$F$+n$ reaction rates: the nucleus $^{20}$Ne is known to be strongly deformed  with strong $\alpha$-cluster structures \cite{ebran2012atomic}, and these structures persist into the neighbouring $^{19}$Ne and $^{19}$F nuclei \cite{PhysRevC.53.1197,PhysRevC.96.044317}. This non-statistical clustering behaviour is not well described by statistical models \cite{PhysRevC.88.015808}. Therefore, experimental values of the resonance properties above the neutron threshold in $^{19}$F are necessary in order to constrain the astrophysical reaction rates.

Direct measurements of neutron-induced reactions on $^{18}$F are made functionally impossible by the difficulty of fashioning either neutrons or $^{18}$F into targets. In the absence of direct measurements, the reaction rates can instead be determined if the properties of the resonances in $^{19}$F above the neutron threshold are known. Some data on neutron-unbound levels in $^{19}$F are available but better constraints particularly on the charged-particle branching ratios are required.

In this paper, we report two experimental studies of excited states in $^{19}$F using in one experiment the Munich Q3D spectrograph, and using in the other experiment the Orsay Enge Split-Pole magnetic spectrograph coupled with an array of silicon detectors. These experiments yield information on the energies and, for resonances for which the widths of the states are larger than the experimental resolution, total widths of the excited levels. The Orsay experiment provides additional information on the relative strength of the charged-particle $p_0$ and $\alpha_0$ decay branches, \textcolor{black}{ giving some  support to the suggestion that the increased production of $^{15}$N is from the $^{18}$F($n,\alpha$)$^{15}$N reaction in helium-burning shell in core-collapse supernovae \cite{PhysRevC.89.025807}. However, it is not yet possible to provide calculated rates for the $^{18}$F($n,p$)$^{18}$O and $^{18}$F($n,\alpha$)$^{15}$N reactions as information on the neutron widths, and the spins and parities of a number of states in $^{19}$F is not available.}

\section{Existing $^{19}$F nuclear data}
\label{sec:19F_existing}

Most existing nuclear data between $E_x = 10.08$ and $10.62$ MeV, corresponding to the region analysed in the present experiment, comes from studies of the resonance reactions \footnote{I.e. these reactions proceed through compound nuclear reactions corresponding to the formation of resonances rather than direct reactions exciting the target nucleus.} with $^{18}$O$+p$ and $^{15}$N$+\alpha$. Some direct reactions have been performed populating states in $^{19}$F via a number of different routes, e.g. $^{18}$F($d,p$)$^{19}$F \cite{PhysRevC.84.054611}, $^{16}$O($^6$Li,$t$)$^{19}$F \cite{PhysRevC.20.1340} etc.; these direct reactions tend to have much poorer energy resolution than the resonance reactions and it is sometimes difficult to firmly identify which states populated in the direct reactions correspond to those observed in resonance reactions. For that reason, the present discussion of existing nuclear data is confined to those data resulting from resonance reactions. A compilation of available nuclear data may be found in Table \ref{tab:Old_19F_levels}. Important aspects of the previous studies are briefly introduced below.

\begin{turnpage}
\begin{table*}[h]
 \caption{Existing $^{19}$F data in the vicinity of the n+$^{18}$F threshold ($S_n=10.4319$~(5)~MeV~\cite{ENSDF}). Energy levels which likely correspond to the levels listed in ENSDF \cite{ENSDF} are listed in the column corresponding to that measurement. The excitation energies from Carlson {\it et al.} \cite{PhysRev.122.607}, Gorodetzky {\it et al.} \cite{GORODETZKY1963462,GORODETZKY1963VERSION2} and Beard {\it et al.} \cite{BEARD1969566} are recalculated using the proton energies listed in those papers and recent mass values \cite{1674-1137-41-3-030003}. The uncertainties are also recalculated, and are dominated by the uncertainties in the proton bombarding energies. Sellin {\it et al.} \cite{SELLIN1969461} and Hesmondhalgh {\it et al.} \cite{HESMONDHALGH1988375} do not report uncertainties on the proton and $\alpha$-particle bombarding energies respectively for observed resonances meaning that uncertainties on the excitation energies may not be given. The excitation energies from the $^{15}$N$+\alpha\rightarrow\gamma$ study of Symons {\it et al.} \cite{0305-4616-4-3-016} are taken directly from that paper. Partial widths are taken from various sources with the references given in each case.}
 \label{tab:Old_19F_levels}

 \begin{tabular}{|c | c | c | c | c | c | c | c | c | c | c | c |}
 \hline
  \makecell{$E_x$ [MeV]\\ \cite{ENSDF}} & $J^\pi$ & \makecell{$\Gamma$ [keV]\\  \cite{ENSDF}} & \makecell{$\Gamma_{\alpha_0}$\\~ [keV]\\\cite{SELLIN1969461}} & \makecell{$\Gamma_{p_0}$\\~ [keV]\\\cite{SELLIN1969461}} & \makecell{$E_x$ [MeV]\\$^{15}$N$+\alpha\rightarrow\gamma$\\\cite{0305-4616-4-3-016}} & \makecell{$E_x$ [MeV]\\$^{15}$N($\alpha,\alpha$)$^{15}$N\\ \cite{HESMONDHALGH1988375}} & \makecell{$E_x$ [MeV]\\$^{18}$O($p,p$)$^{18}$O\\$^{18}$O($p,\alpha$)$^{15}$N \\ \cite{PhysRev.122.607}} & \makecell{$E_x$ [MeV]\\$^{18}$O($p,\alpha$)$^{15}$N \\ \cite{GORODETZKY1963462,GORODETZKY1963VERSION2}} & \makecell{$E_x$ [MeV]\\$^{18}$O($p,p$)$^{18}$O\\$^{18}$O($p,\alpha$)$^{15}$N \\ \cite{SELLIN1969461}} & \makecell{$E_x$ [MeV]\\$^{18}$O($p,n$)$^{18}$F\\ \cite{BEARD1969566}} & Comments \\
  \hline
  \hline
  $10.088(5)$ & $5/2^-, 7/2^-$  & $<1.5$ & & 	& $10.088(5)$ & $10.088$ & 	       & $10.096(6)$	&	& 		& \\
  
  $10.1360(8)$ & $3/2^-$  & $4.3(6)$ & & 	& $10.130(6)$ & 	 & $10.134(7)$ & $10.139(6)$	& 		& 		& \\
  
	& & &	&		&	      & 	 & 	       & $10.155(9)$	&		&		&  \\
	
  $10.162(3)$ & $1/2^+$  & $3.1$ &	$2.2$	& $0.9$ 	&      & 	 & $10.163(7)$ & $10.167(6)$	& $10.161$	& 		& \makecell{$\Gamma = 10$ keV in Refs. \cite{GORODETZKY1963462,GORODETZKY1963VERSION2}.\\ See the note in the text\\ about the width of this state.} \\
  
	&	&	& &	&	      & 	 & 	       & $10.187(8)$	&		&		& \\
	&	&	& &	&	      & 	 & 	       & $10.212(7)$	&		&		& \\		
  $10.231(3)$ & $1/2^+$  & $4.3$ 	& $1.6$ & $2.7$	 & $10.231(3)$ & 	 & $10.246(7)$ & $10.239(6)$	& $10.231$	& 		& \\
  
  $10.253(3)$ & $1/2^+$  & $22.7$ 	& $12.3$ & $10.4$	& 	      & 	 & $10.269(7)$ & $10.264(6)$	& $10.254$	& 		& \\
  
  $10.308(3)$ & $3/2^+$  & $9.2$ & $4.3$ & $4.9$	& $10.308(4)$ & 	 & $10.316(7)$ & $10.315(6)$ 	& $10.308$	& 		& \\
  
  $10.365(4)$ & $7/2, 9/2, 11/2$  & $3.0(15)$ & &	& $10.365(4)$ & 	 & 	       & 		& 		& 		& \\
  
  $10.411(3)$ & $13/2^+$  & $<1.5$ & &	& $10.411(3)$ & $10.411$ &  	       & 		& 		& 		& \\
  
	&     & &          &             &          & $10.427(8)$ & $10.425(10)$  &         	& 	&	& $\Gamma \sim 60$ keV \cite{GORODETZKY1963462,GORODETZKY1963VERSION2} \\
	
  $10.469(4)$ &  & $11.0(12)$ & &	& $10.469(4)$ & 	 & 	       & $10.469(7)$  & 		& 		& \\
  
  $10.488(4)$ &  & $4.8(8)$ & & 	& $10.488(4)$ & 	 & $10.490(8)$ & $10.489(9)$  & 		& 		& \\
  
  $10.4964(10)$ & $3/2^+$ & $4.4$ & $0.9$ & $2.3$ 	& $10.501(4)$ & 	 & $10.501(8)$ & $10.503(7)$  & $10.497$ &  $10.496(1)$ 	& $\Gamma = 10$ keV \cite{GORODETZKY1963462,GORODETZKY1963VERSION2} \\
  
  $10.521(4)$   & $14(2)$ & &	& $10.521(4)$ & 	 & 	       & $10.521(7)$  & 		& 	&	& \\
  
  $10.5423(11)$ & $2.5(2)$  & &	& $10.546(4)$ & 	 & 	       & 		& 	&	& $10.542(1)$ 	& \\
  
  $10.555(3)$ & $3/2^+$ & $4.0(12)$ & &	& $10.554(4)$ & 	 & $10.558(8)$ & $10.555(7)$  & 		& 		&  $T=3/2$\\
  
  $10.5656(11)$ & & $4.6(7)$ & & 	& $10.560(4)$ & 	 	       & 		& 	&	& $10.567(1)$	& \\
  
  $10.580(4)$ & $(5/2^+)$  & $22(3)$ & &	& 	      & 	 & $10.575(8)$ & $10.579(7)$ 	& 		& 		& $\Gamma = 18$ keV \cite{GORODETZKY1963462,GORODETZKY1963VERSION2} \\
  
  &              &         & &      &             &          &             & $10.595(11)$  & 		& 		& \\
  
  $10.6130(16)$ & $5/2^+$ & $5.4$ & $1.1$ & $4.3$	& 	      & 	 & $10.615(8)$ & $10.621(7)$ 	& $10.615$	& $10.614(2)$ 	& $\Gamma = 9$ keV \cite{GORODETZKY1963462,GORODETZKY1963VERSION2}, $T=3/2$ \\
  
   &             &   & &            &             &          &             & $10.676(9)$    &               &               &                \\
  \hline

 \end{tabular}
\end{table*}
\end{turnpage}

The reactions using $^{15}$N$+\alpha$ have been performed with a gas target filled with purified $^{15}$N gas \cite{0305-4616-4-3-016,HESMONDHALGH1988375}. An experiment measuring the yields of the $^{15}$N($\alpha,\gamma$)$^{19}$F, $^{15}$N($\alpha,\alpha^\prime\gamma$)$^{15}$N and $^{15}$N($\alpha,p\gamma$)$^{18}$O reactions by
detection of the associated $\gamma$ rays was performed at Oxford in the 1970s \cite{0305-4616-4-3-016}. In the experiment of Ref. \cite{0305-4616-4-3-016}, nine resonances were reported to be observed in either the $^{15}$N($\alpha,\alpha^\prime\gamma$)$^{15}$N or the $^{15}$N($\alpha,p\gamma$)$^{18}$O channels.

Subsequently, a measurement of the elastic resonant scattering reaction $^{15}$N($\alpha,\alpha$)$^{15}$N was performed at the same facility \cite{HESMONDHALGH1988375}. Only two levels in the region-of-interest were observed in this measurement, at 10.088 and 10.411 MeV. Both of these levels correspond to narrow states in this region.

The resonance reactions involving $^{18}$O$+p$ have been performed with alumina targets enriched in $^{18}$O \cite{GORODETZKY1963462,GORODETZKY1963VERSION2} or thin gas targets \cite{PhysRev.122.607,SELLIN1969461}. Two of the older studies of $^{18}$O$+p$ reactions, those of Carlson {\it et al.} \cite{PhysRev.122.607} and Gorodetzky {\it et al.} \cite{GORODETZKY1963462,GORODETZKY1963VERSION2} use the then-current value for the proton threshold of $^{19}$F, a value which is around 30 keV lower than the present, more accurate value \cite{1674-1137-41-3-030003}. This results in the excitation energies of levels determined in those experiments being around 30 keV below the correct value. The excitation-energy values listed in Table \ref{tab:Old_19F_levels} are re-calculated from the available data in Refs. \cite{PhysRev.122.607,SELLIN1969461,GORODETZKY1963462,GORODETZKY1963VERSION2} using updated mass measurements \cite{1674-1137-41-3-030003}. The uncertainties in the excitation energies are also recalculated - in these cases, they are dominated by the uncertainties in the proton bombarding energies. No information as to the systematic or statistical nature of the uncertainties of the proton bombarding energies in Refs. \cite{PhysRev.122.607,SELLIN1969461,GORODETZKY1963462,GORODETZKY1963VERSION2} are available. This is unfortunate: it is possible that the systematic uncertainties on the proton bombarding energies may be correlated and that the relative uncertainties in the bombarding energies is smaller.

We note that the level observed at $E_x = 10.162(3)$ MeV has a width of $\Gamma = 31$ keV according to Ref. \cite{ENSDF}. However, the corresponding level in Sellin {\it et al.} has a width of $\Gamma = 3.3$ keV \cite{SELLIN1969461} which, if assumed to be the width in the laboratory frame, gives $\Gamma_{cm} = 3.1$ keV, a factor of ten smaller than the width quoted in Ref. \cite{ENSDF}. We assume that the width quoted in Ref. \cite{ENSDF} is the result of a typographical error in Ref. \cite{AJZENBERGSELOVE19721}.

The use of the $^{18}$O($p,n$)$^{18}$F reaction to generate the medical radioisotope $^{18}$F and as a neutron source means that it has been studied in great detail using both activation techniques \cite{hess2001excitation} and direct measurement of the neutron flux \cite{Fritsch1973,BEARD1969566,BAIR1964209,PhysRevC.8.120}. The study of Beard {\it et al.} \cite{BEARD1969566} provides probably the most complete spectroscopy of the $^{18}$O($p,n$)$^{18}$F reaction in the region of the neutron threshold. However, care must be taken when considering the excitation energies determined therein. If modern mass measurements \cite{1674-1137-41-3-030003} are used instead of mass excesses from the compilation in 1960  which is referenced by Beard {\it et al.} \cite{EVERLING1960529}, the excitation energies of the states measured by Beard {\it et al.} are observed to shift by a little less than 2 keV. Bearing this in mind, the excitation energies in Table \ref{tab:Old_19F_levels} have been re-calculated from the proton energies observed by Beard {\it et al.}. The uncertainties in the excitation energies are also recalculated - in these cases, they are dominated by the uncertainties in the proton bombarding energies.

Both the limitation and advantage of resonance reactions is that they are selective in the entrance channel of the populated resonances. Therefore, if the partial width in the entrance channel to a state using a particular reaction is relatively weak the state may drop below the limit-of-detection. A related effect can be observed comparing the study of Sellin {\it et al.} \cite{SELLIN1969461} with those of Gorodetzky {\it et al.} \cite{GORODETZKY1963462,GORODETZKY1963VERSION2} or Carlson {\it et al.} \cite{PhysRev.122.607}: the number of levels observed in the latter two experiments far exceeds the number of those claimed in the former. Careful visual inspection of the spectra of Sellin {\it et al.} \cite{SELLIN1969461} would suggest that some of the states claimed by Gorodetzky and Carlson are, in fact, observed in that experiment but are not treated as such.

In contrast, proton inelastic-scattering reactions at the energies used in the experiments described in this paper are not selective \cite{PhysRevC.89.065805,413,Moss1976429,PhysRevC.89.065805,PhysRevC.102.015801}. Therefore, unlike the previous resonance-reaction experimental studies of $^{19}$F, we should populate most or all of the states present. Using high-resolution, unselective reactions has been used successfully in past experimental studies to help to clarify discrepancies between more selective reaction mechanisms (see, e.g. our previous experimental study of $^{26}$Mg \cite{PhysRevC.97.045807} and references therein).

\section{Munich Q3D}

\subsection{Experiment}

A 16-MeV beam of protons was incident upon a target nominally comprising 40 $\mu$g/cm$^2$ of LiF deposited on a 20-$\mu$g/cm$^2$ natural carbon foil. Scattered protons were momentum-analysed in the Munich Q3D magnetic spectrograph \cite{LOFFLER19731}. The slits at the entrance of the spectrograph were set to 4 mm by 24.5 mm to optimise the energy resolution and to minimise the aberration from contaminating species.

The focal plane consisted of two gas proportional detectors backed by a plastic scintillator. The second proportional detector provides information on the focal-plane position of the detected particle. Particles were identified using the energy losses in the proportional detectors and the remaining energy deposited in the plastic scintillator.

Data were also taken using a $^{28}$SiO$_2$ target (nominally 40 $\mu$g/cm$^2$ on 5 $\mu$g/cm$^2$ carbon) for the purpose both of calibration and quantification of the $^{16}$O background \footnote{Lithium fluoride is hygroscopic and absorbs water readily from the atmosphere.}, and a natural carbon foil (nominally $55$ $\mu$g/cm$^2$) to characterise the background resulting from the $^{12}$C backing of the LiF target.

Data were taken for all targets at 25, 35, 40 and 50 degrees, with two overlapping field settings centred at $E_x = 10.2$ and $10.5$ MeV used to probe the astrophysically important region in $^{19}$F.

\subsection{Data Analysis}

Protons were selected considering the energy losses in both of the proportional detectors and the residual energy detected in the plastic scintillator. The focal plane was calibrated in magnetic rigidity ($B\rho$) at each angle and for each field setting using well-known states in $^{28}$Si.

The background from $^{12}$C was scaled according to the measured charge and the nominal target thicknesses. This was found to under-predict the observed strength of the $9.62$-MeV $J^\pi = 3^-$ state of $^{12}$C in the present experiment. Additional scaling factors were introduced to ensure that the normalisation of the $^{12}$C background was correct for both the LiF and $^{28}$SiO$_2$ targets. This discrepancy is likely due to the nominal thickness of the carbon backing on the targets being inaccurate, potentially due to build-up of carbon residue on the target foil. The carbon background was then subtracted from the experimental spectra. Even following this background subtraction, a significant background is observed in the spectra taken with the LiF target. This background is not present in the spectra taken with the $^{28}$SiO$_2$ target \footnote{For an example $^{28}$SiO$_2$ spectrum taking using the Munich Q3D, refer to Refs. \cite{PhysRevC.97.045807,PhysRevC.102.015801}}. \textcolor{black}{The background likely results from scattering from lithium present within the target and is not instrumental in nature. This is further supported by the relative signal-to-background observed in data taken with a NaF target and a LiF target in a later Munich Q3D experiment which will be the subject of a future publication \cite{MattWilliams} and by the continuous smooth background observed in the coincidence spectra (see Section \ref{sec:OrsayData}). Alternative sources of background include the $^{19}$F($p,\alpha$)$^{16}$O($p$)$^{15}$N reaction and broad states in $^{19}$F above the proton and $\alpha$-particle thresholds. \textcolor{black}{However the latter are excluded as per the discussion in Section \ref{sec:OrsayData}}.}

The location of the contaminating $E_x = 10.356$-MeV state of $^{16}$O ($\Gamma = 26$ keV) was determined by using the $^{28}$SiO$_2$ target. In this case, the spectrum - including any overlapping $^{28}$Si states - was fitted with a linear combination of one exponentially tailed Gaussian function for the $^{16}$O state, and one exponentially tailed Gaussian function for each $^{28}$Si state. This allowed the location and shape of the $^{16}$O state to be described. The contribution of the $^{16}$O state to the LiF focal-plane spectrum was then included using the same function as used to fit the $^{16}$O component of the $^{28}$SiO$_2$ spectrum with a scaling factor to account for the different areal densities of $^{16}$O in the LiF and $^{28}$SiO$_2$ targets.

The resulting $^{19}$F excitation-energy spectra were fitted using a combination of Voigt and exponentially tailed Gaussian functions. The exponentially tailed Gaussian functions were used to describe the slight asymmetry in the Q3D response. The experimental energy resolution, corresponding to the width parameter of the Gaussian function or Gaussian component of the Voigt function, was constrained by the narrow peaks in the spectrum. Both the energy resolution (around $8$ keV FWHM) and the exponential tail parameter (around $1$ keV) were considered identical for all states in each individual fit.

Example spectra along with the resulting fits to the data from two different field settings and two different angles are shown in Figures \ref{fig:Q3DSpectra35Deg} and \ref{fig:Q3DSpectra40Deg}. The centroids of observed states are shown by vertical lines. Individual contributions from states are also included in the figures, see the captions for details.

\begin{figure*}[htbp]
 \includegraphics[width=\textwidth]{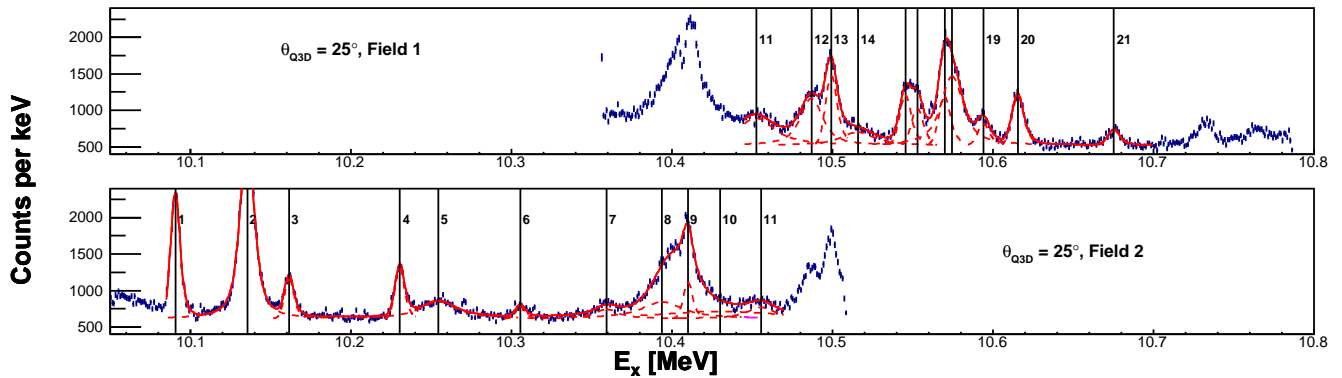}
 \caption{Excitation-energy spectra resulting from the $^{19}$F($p,p^\prime$)$^{19}$F reaction at $\theta_{\mathrm{Q3D}} = 25^{\circ}$. The total fit is shown as a solid red line and individual state contributions are shown as dashed red lines. The $10.345$-MeV state in $^{16}$O is shown as a solid purple curve. Vertical solid black lines show the locations of states with the associated indices corresponding to those listed in Table \ref{tab:Q3DResults}. The two different fields correspond to different strengths of the magnetic fields of the Q3D to centre two different excitation energies on the focal plane of the spectrograph. Note that the y-axis for this spectrum does not start at 0 and so the extent of the uniform background on the focal plane is somewhat hidden.}
 \label{fig:Q3DSpectra25Deg}
\end{figure*}

\begin{figure*}[htbp]
 \includegraphics[width=\textwidth]{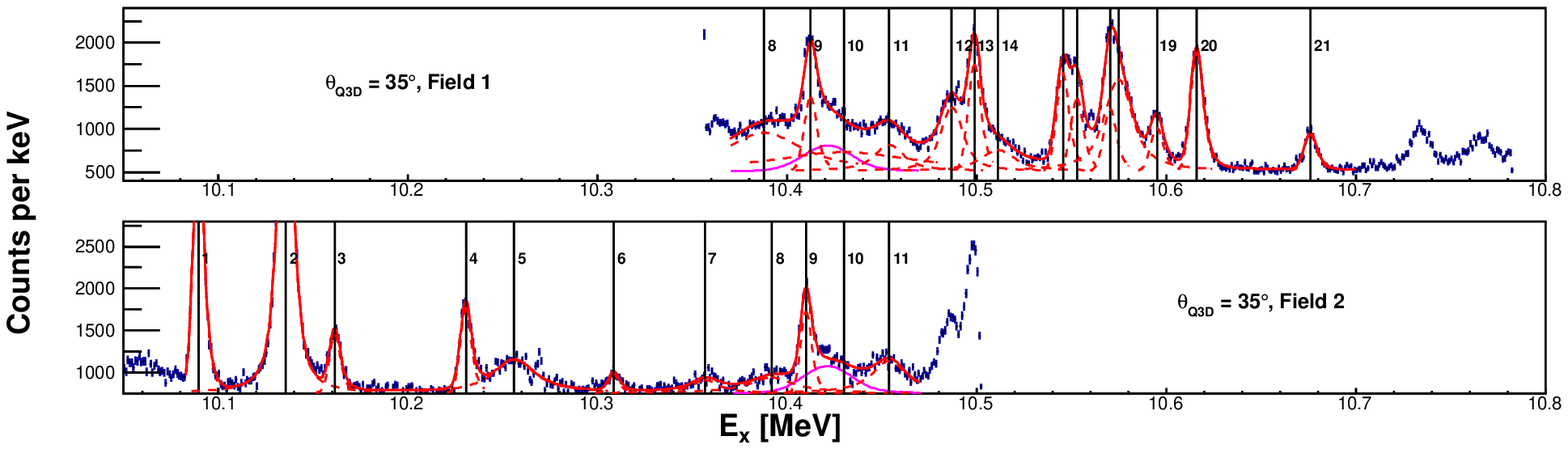}
 \caption{As Figure \ref{fig:Q3DSpectra25Deg} but for data at  $\theta_{\mathrm{Q3D}} = 35^{\circ}$.}
 \label{fig:Q3DSpectra35Deg}
\end{figure*}

\begin{figure*}[htbp]
 \includegraphics[width=\textwidth]{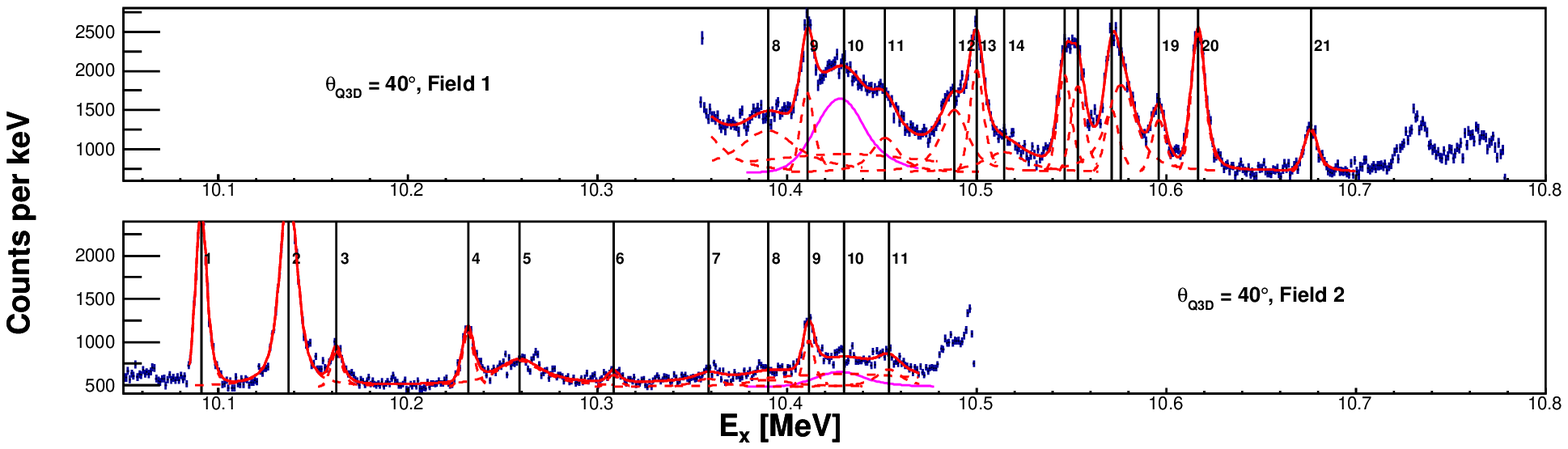}
 \caption{As Figure \ref{fig:Q3DSpectra25Deg} but for data at  $\theta_{\mathrm{Q3D}} = 40^{\circ}$.}
 \label{fig:Q3DSpectra40Deg}
\end{figure*}

\begin{figure*}[htbp]
 \includegraphics[width=\textwidth]{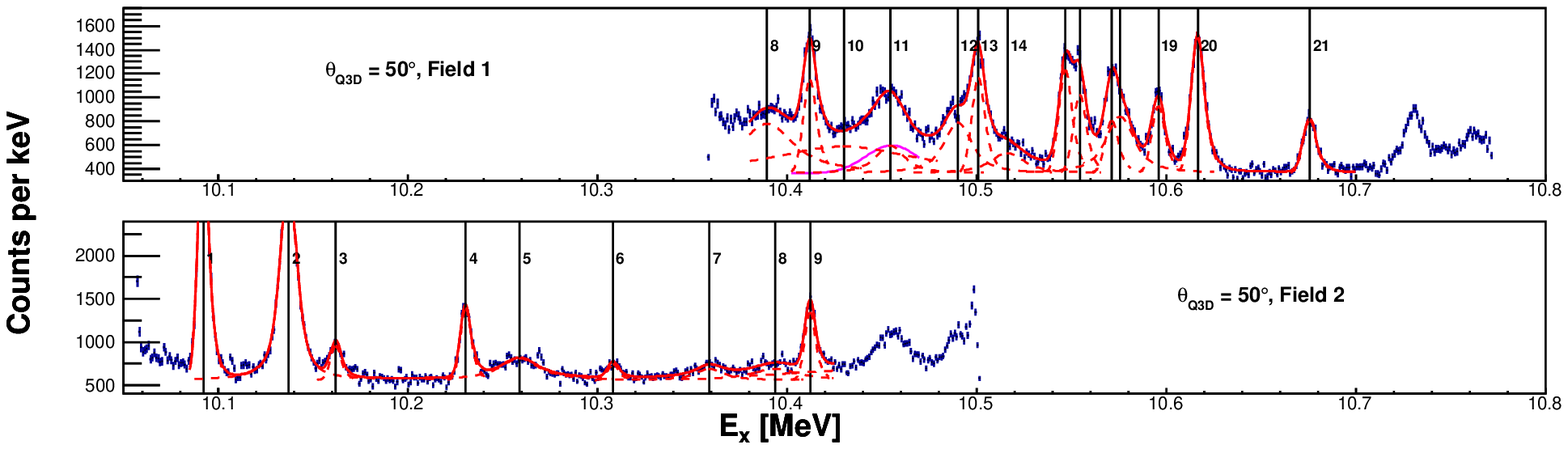}
 \caption{As Figure \ref{fig:Q3DSpectra25Deg} but for data at  $\theta_{\mathrm{Q3D}} = 50^{\circ}$.}
 \label{fig:Q3DSpectra50Deg}
\end{figure*}


\subsection{Results}

The parameters of the levels extracted from the Munich data are given in Table \ref{tab:Q3DResults}. The uncertainties on the excitation energies of the levels is purely statistical, and is taken from the weighted average of the excitation energies determined at each angle. \textcolor{black}{The uncertainties included an additional contribution to account for unknown systematic effects to ensure that the reduced $\chi^2$ of the state parameters at each angle were no greater than 1 using the method described in Ref. \cite{PhysRevC.85.065809}.}

\begin{table*}
\caption{Energy levels determined from the Munich Q3D experiment. The index of the state is given for ease of reference during the discussion in the text. A \textquoteleft narrow\textquoteright\ state is one which was fitted with an exponentially tailed Gaussian function and not a Voigt function. Where possible, a suggested correspondence has been made between levels observed in the present experiment and those from previous experimental studies as summarised in Table \ref{tab:Old_19F_levels}. All excitation-energy uncertainties from the present experiment reported in this table are purely statistical. The origin and magnitude of possible systematic errors is described in text.}
\label{tab:Q3DResults}
\begin{ruledtabular}

 \begin{tabular}{ c  c  c  c }

  State Index & $E_x$ [MeV] & $\Gamma$ [keV] & \makecell{Suggested corresponding level \\ from Table \ref{tab:Old_19F_levels} [MeV]} \\
  \hline
  1  & $10.091(1)$ & Narrow & $10.088(5)$ \\
  2  & $10.136(1)$ & $6(1)$ & $10.1360(8)$ \\
  3  & $10.162(1)$ & Narrow & $10.162(3)$ \\
  4  & $10.231(1)$ & Narrow & $10.231(3)$ \\
  5  & $10.257(2)$ & $25(2)$ & $10.253(3)$ \\
  6  & $10.308(1)$ & Narrow & $10.308(3)$ \\
  7  & $10.359(2)$ & $15(4)$ & $10.365(4)$ \\
  8  & $10.392(2)$ & $29(7)$ & \\
  9 & $10.411(1)$ & Narrow & $10.411(3)$ \\
  10 & $10.420(10)$ & $>60$ & \\
  11 & $10.453(2)$ & $25(3)$ & \\
  12 & $10.488(2)$ & $12(2)$ & $10.488(4)$ \\
  13 & $10.500(1)$ & $11(2)$ & $10.4964(10)$ \\
  14 & $10.515(2)$ & $19(3)$ & $10.521(4)$ \\
  15 & $10.546(2)$ & Narrow & $10.5434(11)$ \\
  16 & $10.554(2)$ & Narrow & $10.555(3)$ \\
  17 & $10.575(2)$ & Narrow &  \\
  18 & $10.579(2)$ & $14(2)$ & $10.580(4)$ \\
  19 & $10.595(1)$ & Narrow & $10.595(11)$ Refs. \cite{GORODETZKY1963462,GORODETZKY1963VERSION2} \\
  20 & $10.616(1)$ & Narrow & $10.6130(16)$ \\
  21 & $10.676(1)$ & Narrow & $10.676(9)$ Refs. \cite{GORODETZKY1963462,GORODETZKY1963VERSION2} \\
 \end{tabular}
 \end{ruledtabular}

\end{table*}

A complete uncertainty budget is given in Table \ref{tab:Q3DUncertaintyAnalysis}. When comparing the differences in the excitation energies determined in the present experiment from those from previous experimental studies, we find that the deviations observed are within the experimental uncertainties. However, this may merely reflect the dominance of systematic uncertainties in previous experimental studies.

The angular uncertainty is assumed to originate in the read-off of the angle setting for the Q3D, which is in gradations of 0.1\textdegree. However, the calibration is performed assuming scattering at a particular angle to calculate the proton rigidities. Therefore, the effect of the angular uncertainty on the excitation-energy uncertainty is relatively small, and is caused by the differing kinematic shifts of the $^{19}$F($p,p^\prime$) and $^{28}$Si($p,p^\prime$) reactions.

The calibration uncertainty is determined by propagating the uncertainties extracted on each of the parameters of the quadratic conversion from the focal-plane position to magnetic rigidity and thence to excitation energy. This contribution to the uncertainty budget does not include possible energy-loss or target-thickness effects on the calibration, which are accounted for separately.

The target-thickness and energy-loss contributions to the uncertainty are both assumed here as fractional contributions relative to the nominal thickness. The fractional contribution of the target thicknesses was taken from the scaling that had to be applied to the $^{12}$C background in order to match the experimental data. As the targets are extremely thin, the energy loss for the protons through the foils is typically between 1 and 2.5 keV, and the corresponding uncertainties introduced by the target thickness or the energy losses are negligible.

The effects of field variations were determined in the same manner as described in Ref. \cite{PhysRevC.97.045807} -  two isolated narrow states at $E_x = 10.088$ and $E_x = 10.616$ MeV were fitted for sections of different runs and the shifts in the peak positions were measured. Shifts of a little less than 1 keV were observed. The shifts were within the fitting uncertainties for the excitation energies.

\begin{table}[htpb]
\caption{Uncertainty budget for the Munich Q3D experiment. For the origin of each contribution of the uncertainty, see the text. The total combined uncertainty is the sum in quadrature of the individual components.}
\label{tab:Q3DUncertaintyAnalysis}

\begin{ruledtabular}
 \begin{tabular}{ c  c  c }
 Origin & Magnitude & Contribution (keV) \\
 \hline
 Angle & 0.1\textdegree & 0.8 \\
 Calibration & From fit & 1.2  \\
 Target thickness & 20\% & 0.4  \\
 Energy losses & 10\% & 0.2 \\
 Field variations & From data & 1.0 \\
 \hline
 Total & & 2 \\
 \end{tabular}
 \end{ruledtabular}

\end{table}

\subsection{Discussion}

Most of the levels observed in the present experiment have a corresponding observed level from the resonant reactions listed in Table \ref{tab:Old_19F_levels}. For completeness, we discuss cases where the correspondence between levels listed in the ENSDF \cite{ENSDF} and those observed in the present experiment is unclear, or where the status of a state is uncertain.

Some states listed in Table \ref{tab:Old_19F_levels} at $E_x = 10.469$ and $10.5656$ MeV are not used in the fitting of the Q3D data. This could be because these states are not populated in the present reaction, which is unlikely given the non-selective nature of the ($p,p^\prime$) reaction at low energies \cite{PhysRevC.102.015801}, or because \textcolor{black}{these states correspond to other known states which have been reported in different reactions. Other factors, such as revisions to the masses (and therefore particle thresholds), hinder the consistent identification of states between different experiments and may help to explain the discrepancies in the number and energy of the reported states.}

\subsubsection{8: The 10.392-MeV state}

A new state is observed at $E_x = 10.392(2)$ MeV at all measured angles. The width is measured to be $\Gamma = 39(4)$ keV. There is no obvious corresponding state observed in any other experiments. It is possible that this state has been missed in resonant-scattering experiments due to its large width 

\subsubsection{10: The 10.420-MeV state}

In the $^{18}$O($p,\alpha$)$^{15}$N experiments of Gorodetzky \cite{GORODETZKY1963462,GORODETZKY1963VERSION2} and Carlson \cite{PhysRev.122.607}, a broad ($\Gamma \sim 60$ keV) state was observed at $E_x = 10.426(10)$ MeV. This state has typically been omitted in compilations of the levels of $^{19}$F \cite{ENSDF}.

In order to ascertain whether this state is real we performed the analysis with and without the state included. No significant improvement was observed in the quality of the fits to the spectra for the Munich Q3D measurement. However, this state was found to be necessary to describe the $\alpha_0$ coincidence spectrum (see Section \ref{sec:OrsayData}) and so it was included in the fits of the Q3D with parameters of $E_x = 10.420(10)$ MeV and $\Gamma = 90$ keV. The Q3D data are not able to provide significant constraints for the properties of this state since the state is broad and there is a continuum background.

\subsubsection{11: The 10.452-MeV state}

A new level is observed at $E_x = 10.452(2)$ MeV with $\Gamma = 25(3)$ keV. This state can only be observed in the $\theta_{Q3D} = 25^\circ$ and $35^\circ$ data; at $40^\circ$ and $50^\circ$ it is obscured by the $^{16}$O background.

The only previously observed level at around this excitation energy is the broad ($\Gamma \sim 60$ keV) state at $E_x = 10.426$ MeV discussed above. As both the excitation energy and width of these states are so different, it is unlikely that these states are the same. We conclude that this is a previously unobserved state in $^{19}$F.

\subsubsection{14: The 10.516-MeV state}

The state at $E_x = 10.516(2)$ MeV is newly observed in the present experiment with a width of $\Gamma = 22(3)$ keV. It is only populated weakly and lies just above the strongly populated state at $E_x = 10.500(1)$ MeV.

A state at $E_x = 10.521(7)$ MeV was observed in Refs. \cite{GORODETZKY1963462,GORODETZKY1963VERSION2,0305-4616-4-3-016}. The resonance observed in the $^{18}$O($p,\alpha$)$^{15}$N study of Gorodetzky {\it et al.} \cite{GORODETZKY1963462,GORODETZKY1963VERSION2} is extremely weak which may explain why this state was not observed in the $^{18}$O$+p$ measurement of Carlson {\it et al.} \cite{PhysRev.122.607} or Sellin {\it et al.} \cite{SELLIN1969461}. No width is quoted for this state in those reference.

A state at $E_x = 10.521(4)$ MeV ($\Gamma = 14(2)$ keV) was observed in Ref. \cite{0305-4616-4-3-016}. This state was observed in $^{15}$N$+\alpha\rightarrow\gamma$ reactions, probably in the $^{15}$N($\alpha_0,\alpha_1$)$^{15}$N channel though Ref. \cite{0305-4616-4-3-016} does not explicitly state this.

It is not clear if the state observed in the present experiment corresponds to those observed in previous measurements.

\subsubsection{17: The 10.571-MeV state}
A new level is observed at $E_x = 10.571(2)$ MeV. This state is not resolved from the broader level at $E_x = 10.579(2)$ MeV. Only using one state at $E_x = 10.57-10.58$ MeV results in a poor fit at all angles.

\subsubsection{18: The 10.579-MeV state}

A state has been observed in the $^{18}$O$+p$ studies of Prosser {\it et al.} \cite{PhysRev.157.779}, Carlson {\it et al.} \cite{PhysRev.122.607} and Gorodetzky {\it et al.} \cite{GORODETZKY1963462,GORODETZKY1963VERSION2} at $E_x = 10.580(4)$ MeV with $\Gamma = 22(3)$ keV \cite{PhysRev.157.779} or $\Gamma = 18$ keV \cite{GORODETZKY1963462,GORODETZKY1963VERSION2}. In the present experiment, there is a state observed at this excitation energy but with a width of $\Gamma = 11(2)$ keV, significantly smaller than the width in the ENSDF database \cite{ENSDF}.

The width in the ENSDF database comes from a study of the $^{18}$O($p,n$)$^{18}$F, $^{18}$O($p,p^\prime\gamma$)$^{18}$O and $^{18}$O($p,\alpha_{1,2}\gamma$)$^{15}$N reactions by Prosser {\it et al.} \cite{PhysRev.157.779}. The widths observed in that measurement are typically much higher than the widths observed in the measurement of Beard {\it et al.} \cite{BEARD1969566} and it is plausible that the widths in Ref. \cite{PhysRev.157.779} are systematically overestimated. For this reason, we conclude that the resonance observed in Prosser {\it et al.} is the same as the resonance observed in the present measurement, and has a width of $\Gamma = 11(2)$ keV.

\subsubsection{21: The 10.676-MeV state}

The state at $E_x = 10.676(1)$ MeV is not listed in current nuclear data compilations \cite{ENSDF}. This state may correspond to the resonance observed at $E_p = 2831(9)$ keV ($E_x = 10.676(9)$ MeV) in the $^{18}$O($p,\alpha_0$)$^{15}$N study of Gorodetzky {\it et al.} \cite{GORODETZKY1963462,GORODETZKY1963VERSION2}. In Carlson {\it et al.} \cite{PhysRev.122.607}, only one resonance is observed ($E_p = 2824(8)$ keV) compared to the two in Gorodetzky {\it et al.} at $E_p = 2815(9)$ and $2831(9)$ keV. It is possible that the resonance listed in Carlson is the unresolved strength of the two resonances observed by Gorodetzky {\it et al.} \cite{GORODETZKY1963462,GORODETZKY1963VERSION2}.

\section{Orsay Split-Pole}

\subsection{Experiment}

A beam of 15-MeV protons was incident upon a target comprising 84 $\mu$g/cm$^2$ of LiF deposited on a 32-$\mu$g/cm$^2$-thick $^{12}$C foil. Scattered protons were momentum-analysed in an Enge Split-Pole magnetic spectrometer. The aperture of the spectrometer covered 1.3 msr and was placed at $\theta_{\mathrm{lab}} = 30^{\circ}$ and $40^{\circ}$. Data were taken at two angles so that contaminants could be identified but coincidence data with the silicon detectors (see below) were only taken at $\theta_{\mathrm{lab}} = 40^{\circ}$. The focal plane consisted of a position-sensitive gas detector backed by a gas proportional detector and a plastic scintillator. Focal-plane particle identification was accomplished using the energy deposition in the gas proportional detector and the focal-plane position.

An array of 6 silicon detectors (\textquoteleft W1\textquoteright\ design from Micron Semiconductor Ltd. \cite{W1}) was placed within the scattering chamber of the spectrometer. The spectrometers were placed at backward angles at around 110 mm (detectors 1-4) or 90 mm (detectors 5 and 6) from the target. Detectors 1 and 2 covered $110<\theta_\mathrm{lab}<125$ degrees, detectors 3 and 4 covered $135<\theta_\mathrm{lab}<165$ degrees, and detectors 5 and 6 covered $110<\theta_\mathrm{lab}<150$ degrees. Detectors 1-4 were placed to the right of the beam and detectors 5 and 6 to the left. Charged-particle decays resulting from inelastic scattering reactions were detected in the silicon array. The signals from the silicon detectors were amplified in Mesytec MPR preamplifier modules, the signals were then transmitted to Mesytec STM-16 leading-edge discriminator modules which gave a shaped output energy signal and an ECL timing signal. Energy signals were recorded for strips on the junction and Ohmic sides of each detector. Timing signals were recorded for junction sides only.

The trigger for the experiment was a coincidence between the gas proportional detector and the plastic scintillator at the focal plane of the Enge spectrometer. The shaping time of the STM-16 amplifiers was set so that the silicon energy signal fell after the trigger from the focal plane-detectors. The timing window of the Caen V1190A time-to-digital converters was set so that it included the timing signals from the silicon detectors which precede the trigger from the spectrometer focal plane. 

\subsection{Data Analysis\label{sec:OrsayData}}

Inelastically scattered protons detected at the focal plane were selected using the energy deposited in the gas proportional detector and the focal-plane position. The focal-plane spectrum was then converted to magnetic rigidity, $B\rho$, using a magnetic field which was logged during the experiment, correcting any variations in the focal-plane position caused by shifts in the magnetic field. The calibration of the focal plane is made by considering the energies of known levels in $^{19}$F as determined from the experiment performed using the Munich Q3D.

Silicon hits were accepted if the energy deposited in the front and back of a detector was within 80 keV, and the time between the silicon and focal-plane events fell within a given kinematic locus. Two-dimensional matrices of the missing energy against the excitation energy were constructed for valid silicon events, assuming a particular reaction channel. The missing energy is the deficit between the known initial energy and the sum of the final energies of the reaction products, requiring an assumption about the kinematics of the reaction. For a reaction of beam $a$ on target $A$ resulting in the ejectile $b$ and heavy recoil $B$ which subsequently decays into light fragment $c$ and heavy fragment $C$, the missing energy $M$ is:

\begin{equation}
\begin{split}
M = T_a - T_b - T_c - T_C.
\end{split}
\end{equation}
$T_a$, the kinetic energy of the beam, is defined by the accelerator, $T_b$ is the ejectile proton energy measured in the Split-Pole, $T_c$ is the energy of the light decay particle measured in the silicon detectors and $T_C$ is the kinetic energy of the heavy fragment and is calculated from the kinematic information of the beam, ejectile and light decay fragment under the requirement that the 4-momentum is conserved.

\begin{figure*}[htbp]
\includegraphics[width=\textwidth]{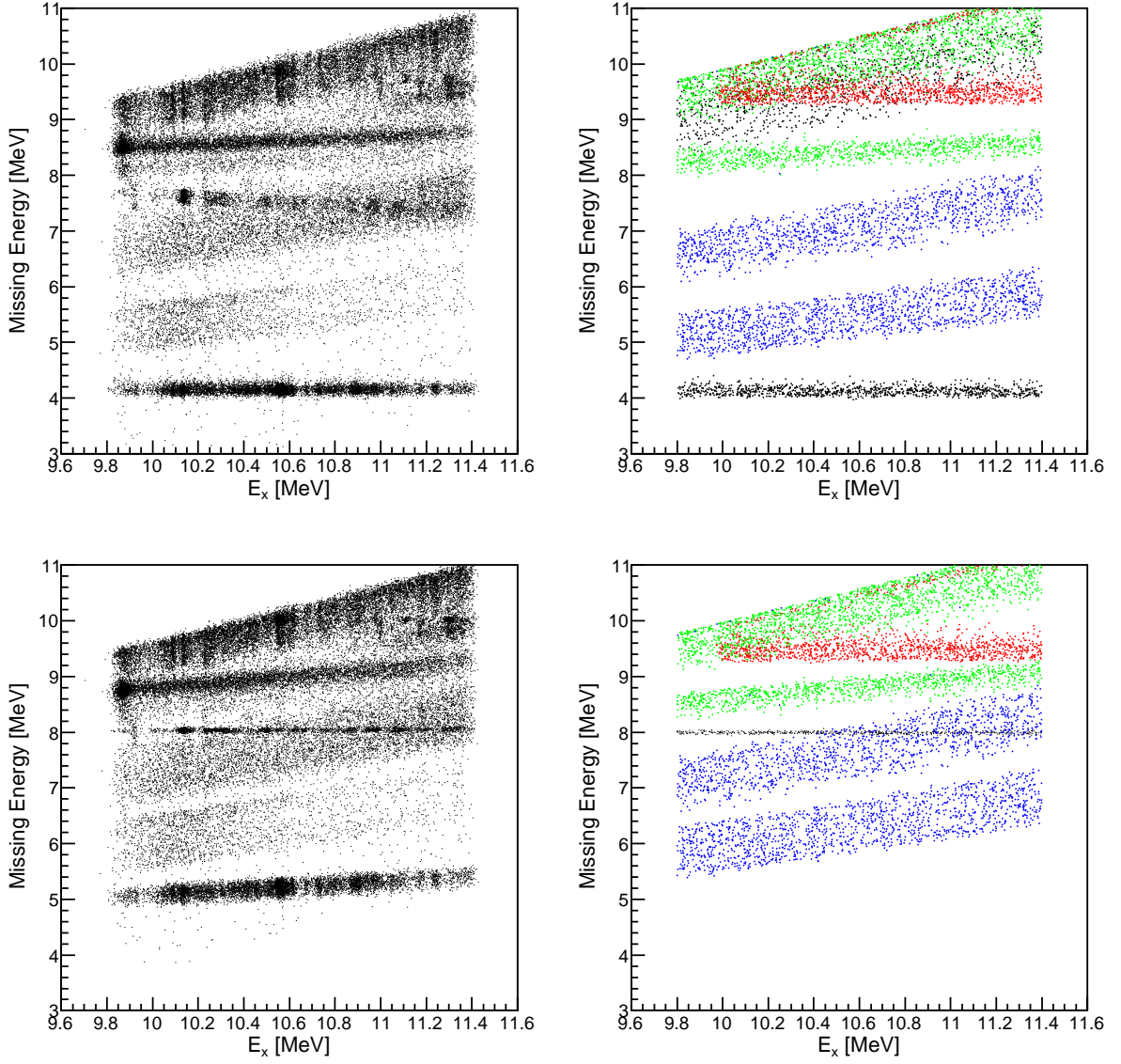}
 \caption{Matrices of experimental and simulated coincidence events. The excitation energy (abscissa) is plotted against the missing energy (ordinate) for DSSSD detectors 1 and 2 covering the same angular range. (Top left) The experimental coincidence matrix assuming $^{19}$F($p,p^\prime$)$^{19}$F($\alpha$)$^{15}$N reaction kinematics. (Top right) The simulated coincidence matrix assuming $^{19}$F($p,p^\prime$)$^{19}$F($\alpha$)$^{15}$N reaction kinematics for various different reactions, including (black) $^{19}$F with $\alpha$ decays to the ground state of $^{15}$N, (red) $^{19}$F with $\alpha$ decays to the first-excited $E_x = 5.27$-MeV state in $^{15}$N, (green) $^{12}$C decays through the ground state of $^8$Be and (blue) $^7$Li decay into $\alpha$+t. (Bottom left) The experimental coincidence matrix assuming the $^{19}$F($p,p^\prime$)$^{19}$F($p$)$^{18}$O) reaction kinematics. (Bottom right) As top right but assuming $^{19}$F($p,p^\prime$)$^{19}$F($p$)$^{18}$O reaction kinematics. The locus at high values of missing energy is due to the detection of $^{15}$N ions and low energy $\alpha$ particles from the decay of $^8$Be.}
 \label{fig:2DMatrices}
\end{figure*}

The identification of different reaction channels and reactions from target contaminants is simpler using the missing energy. An example of two of these two-dimensional matrices assuming $\alpha$-particle and proton decays from states in $^{19}$F are shown in Figure \ref{fig:2DMatrices}. Decay channels from $^{19}$F states with the correct kinematic reconstruction appear as horizontal loci. Other reaction channels and target contaminants have sloped loci. Gates can then be placed on the loci corresponding to different reaction channels.

\textcolor{black}{Since the target is lithium fluoride on a carbon backing, the expected reactions include reactions from $^{6,7}$Li and $^{19}$F as well as $^{12}$C from the carbon backing and $^{16}$O from water absorbed by the target. Simulations of the $(p,p^\prime$) reactions with subsequent proton and $\alpha$-particle decays were performed with GEANT4. The contribution from $^6$Li to the coincidences is small and is omitted from Figure \ref{fig:2DMatrices}. One main contamination is from the decay of $^7$Li into $\alpha+^3\mathrm{H}$ following excitation which is the cause of two of contaminating loci. The other main contaminating channel is the decay of $^{12}$C into 3 $\alpha$ particles via the ground state of $^8$Be. The simulations of this locus slightly underpredict the missing energy; the carbon is predominantly from the backing of the target and the $\alpha$ particles have to traverse the entire target to reach the DSSSDs, resulting in a lower kinetic energy detected in the DSSSDs and therefore a mild increase in the missing energy. The $^{16}$O($p,p^\prime$)$^{16}$O($\alpha$ coincidence locus is omitted from Figure \ref{fig:2DMatrices} since $^{16}$O has a peak in the region of interest and not a significant continuous distribution.}

\textcolor{black}{The coincidence locus from $^7$Li in the experimental data is rather smooth across the entire excitation-energy region of interest, supporting our suggestion that the smooth background is due to reactions from the lithium in the target. Simulations performed of the $^{19}$F($p,\alpha$)$^{16}$O($p$)$^{15}$N reaction channel for the Orsay data predict additional loci should be present in the coincidence spectra from that experiment (again, see Section \ref{sec:OrsayData}). Since those loci are not observed, we do not expect the proton background from this reaction channel to contribute to the focal-plane spectra. Broad $^{19}$F states would have to decay by $\alpha$-particle or proton emission and would appear strongly in the coincidence spectra shown in Section \ref{sec:OrsayData}; since they do not, the focal-plane background cannot result from broad states in $^{19}$F.
}

\begin{figure}[htbp]
    \centering
    \includegraphics[width=0.45\textwidth]{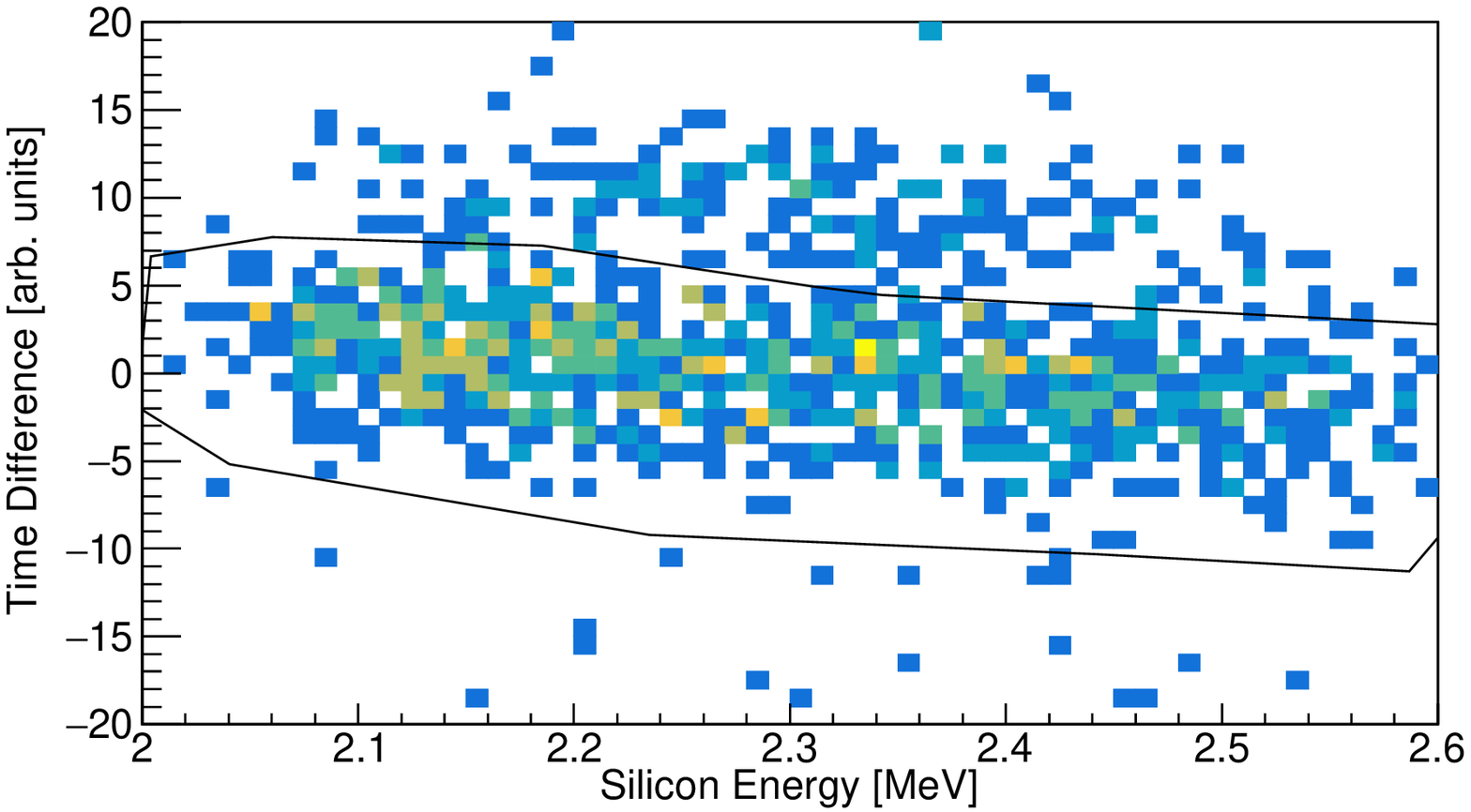}
    \includegraphics[width=0.45\textwidth]{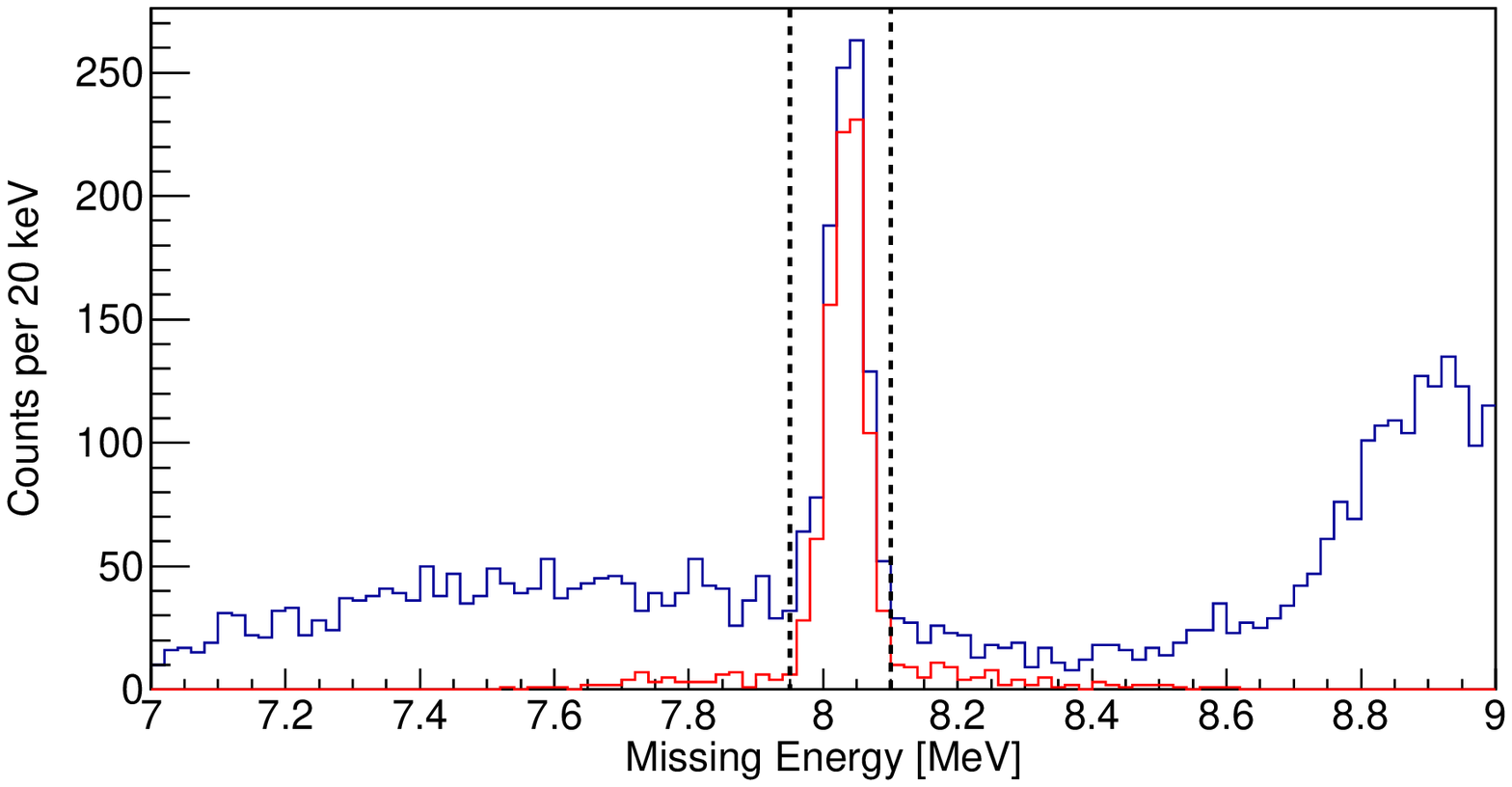}
    \caption{(Top) Energy of the particle detected in the silicon detector (abscissa) against the time difference between the focal-plane event and the particle detected in the silicon detector (ordinate). The additional timing gate on the proton events is also shown. This spectrum was generated using events from detectors 1 and 2 with a gate on $E_{\mathrm{x}}$ between 10.2 and 10.7 MeV and on values of the missing energy (assuming proton kinematics) between 7.95 and 8.1 MeV.
    (Bottom) Missing energy spectra with a gate on time differences from -20 to 20 (blue) and with the more constrictive timing gate (red). The suppression of the events from the contaminant channels is clear. The dotted vertical lines in the bottom panel show the gates on the missing energy which are used to generate the time-difference spectrum in the upper panel.}
    \label{fig:p0_selection}
\end{figure}

\begin{figure*}[htbp]
\includegraphics[width=0.95\textwidth]{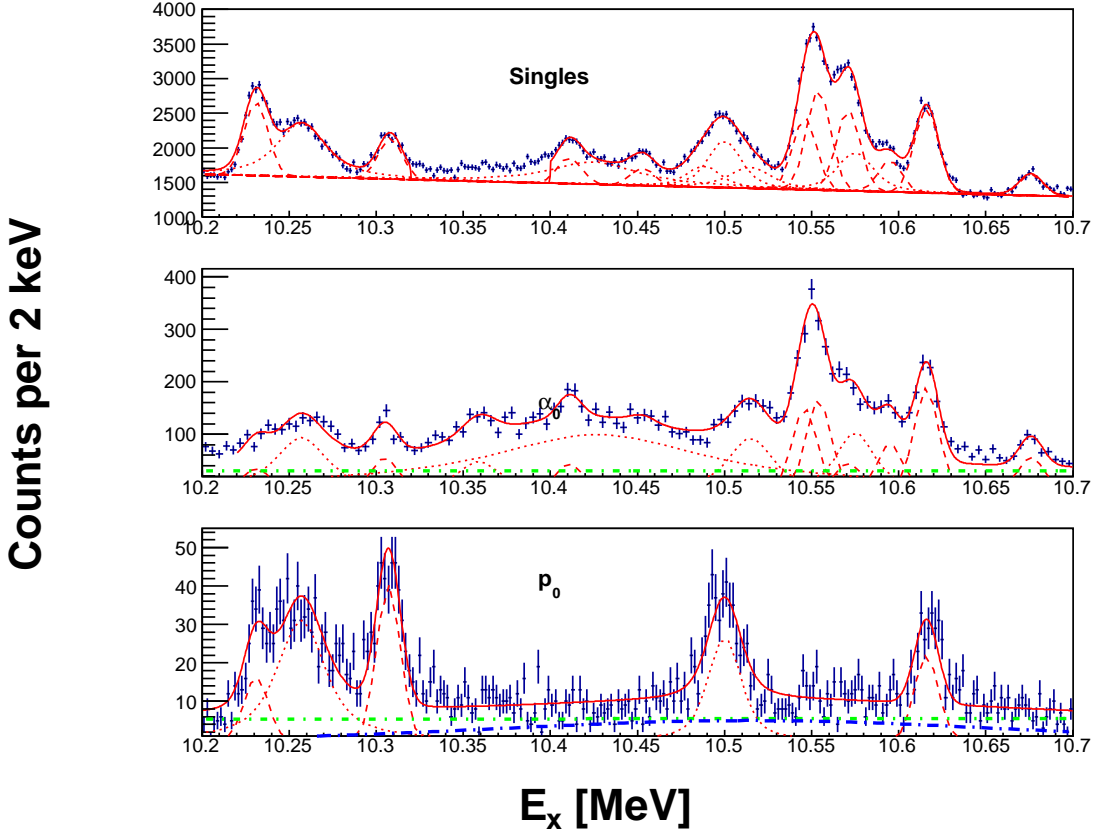}
 \caption{Focal-plane spectra from the Orsay experiment. Top: Singles (inclusive) spectrum. Middle: $\alpha_0$-gated focal-plane spectrum. Bottom: $p_0$-gated focal-plane spectrum. The solid lines show the overall fits of the spectra. The dotted and dashed lines show the contributions from individual resonances. The green and blue additional lines in the $\alpha_0$ and $p_0$ show the linear background component and the additional background component from the contaminant channels in the $p_0$ locus in detectors 5 and 6. Resonances which are not shown in the $p_0$ spectrum are consistent with no signal and instead the maximum yield is computed as an upper limits; see the text for additional details.}
 \label{fig:OrsaySpectra}
\end{figure*}

The $\alpha_0$ locus is very clean. This is because of the low $\alpha$-particle threshold in $^{19}$F: the $\alpha$-particles decaying from states in the excitation-energy range populated in the proton-scattering reaction are much higher in energy than those from reaction channels resulting from contaminating nuclei. The $p_0$ channel is, however, embedded in a region with a number of contaminating channels. The effect of these channels may be reduced/removed using tighter timing gates for detector 1 to 4; an example of this is shown in Figure \ref{fig:p0_selection}.  For detectors 5 and 6, which are closer to the target, this is not possible and the background is instead included in the fit of the data. This background is determined by the linear interpolation of the backgrounds found with gates on the missing energy. The generated inelastic proton spectra in coincidence with $\alpha_0$ and p$_0$ as well as the inclusive (singles) spectrum are shown in Figure \ref{fig:OrsaySpectra}.

The parameters ($E_x, \Gamma$) of states determined from the Munich Q3D data were used to fix the relevant fitting parameters for the Orsay Split-Pole data with the exception of the $E_x = 10.420(10)$-MeV broad state. The parameters for this state were determined from the combined $\alpha_0$ spectrum as it provides the cleanest signal from the broad state; the width of this state is estimated to be $\Gamma = 105(30)$ keV with the continuum background again making estimation of the parameters of this resonance difficult. The exclusive spectra were fitted using a combination of Gaussian and Voigt functions. The experimental resolution ($16$ keV FWHM) was fixed using narrow states in the $p_0$ decay locus.

In the case of the $p_0$ decay data (Figure \ref{fig:2DMatrices}, right panel), there is an overlapping locus resulting from other reaction channels. To account for this background in the fitting of the data, the background is described by a Gaussian function. The centroid and width of the background component are determined by generating the coincidence spectra gating above and below the $p_0$-decay locus in the missing energy vs excitation energy plots. The centroids and widths of the background function are then determined by linear interpolation of the centroids and widths of the background components generated from the off-$p_0$-decay loci.

Branching ratios of excited states were determined by comparing the yields of state in the coincidence spectra to the yields of states in the inclusive spectrum. To account for anisotropy in the decays, the yields for various angular regions in the silicon detectors were calculated using the GEANT4 simulation for isotropic decays and decays with an angular correlation function following $W(\theta) = 1 + P_2(\mathrm{cos}(\theta))$ and $W(\theta) = 1 + P_2(\mathrm{cos}(\theta)) + P_4(\mathrm{cos}(\theta))$ where $P_2$ and $P_4$ are the second- and fourth-order Legendre polynomials, respectively. Here, $\theta$ is the angle of the decaying particle relative to the motion of the $^{19}$F recoil.

The yields for each state were then calculated assuming that the angular correlation function is:

\begin{equation}
    W(\theta) = \sum_{k=0}^2 A_k P_{2k}(\mathrm{cos}(\theta)),
\end{equation}
where the $A_i$ are real coefficients found by minimising the $\chi$-squared function.

\begin{figure}
    \centering
    \includegraphics[width=\columnwidth]{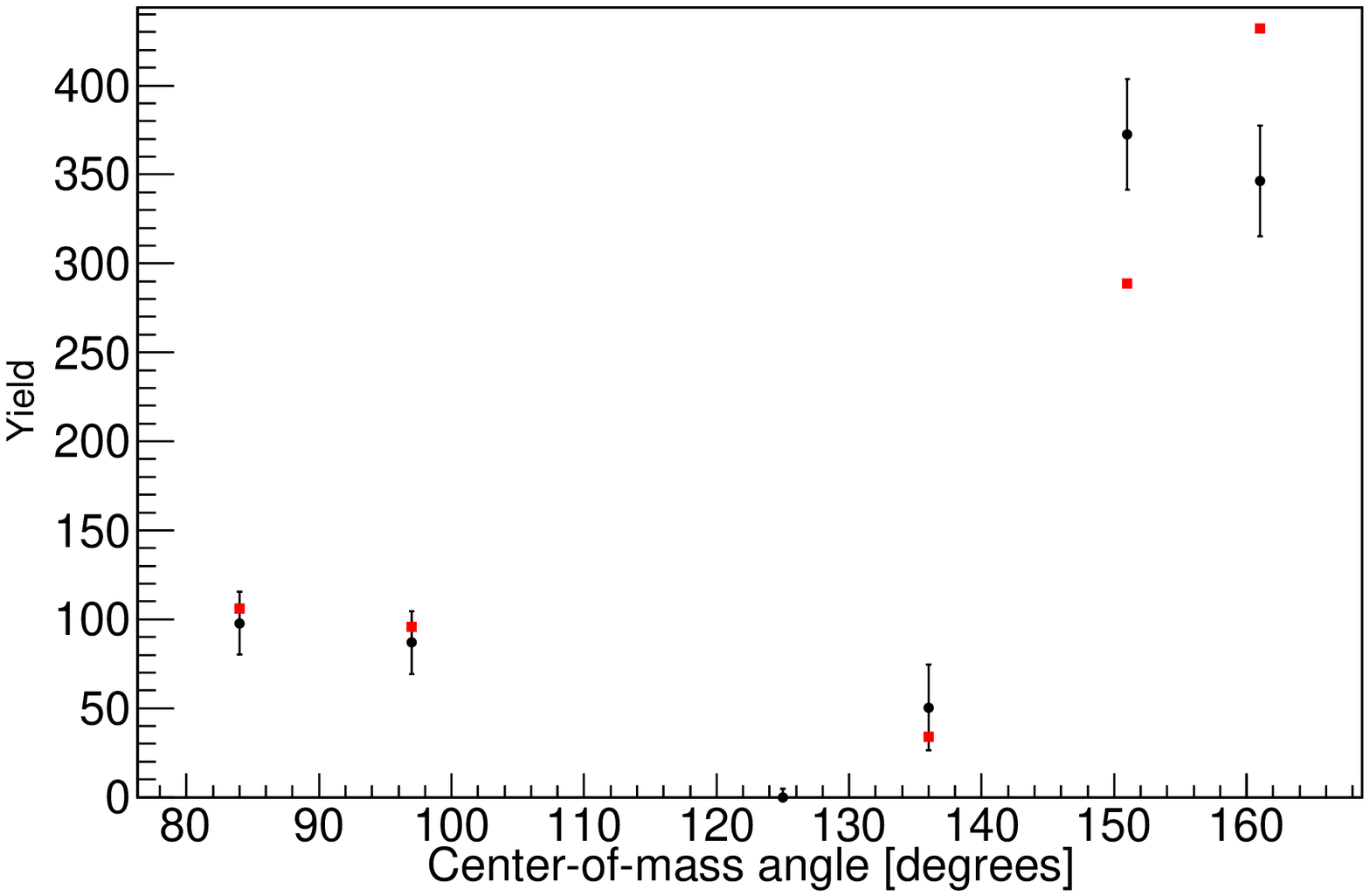}
    \caption{The (black) experimental and (red) fitted yield per region for the state at $E_x = 10.256$ MeV.}
    \label{fig:AngularCorrelationExample}
\end{figure}

The total yield is extracted from the integration of the angular correlation function. The procedure to extract the yields was tested with a simulated dataset using the angular correlation function extracted for one of the observed states. The yields per detector and the total yield used in the simulation were successfully reproduced showing that the method used to extract the yields is robust. An example experimental angular distribution with fit is shown in Figure \ref{fig:AngularCorrelationExample} for the $\alpha_0$ decay from the $E_x = 10.256$-MeV state. The uncertainty on the yield comes from the uncertainty in the coefficient of the 0th-order polynomial as all other terms in the expansion cancel in the integration over the polar angle.

Due to the high level of background in the inclusive spectrum from the Orsay measurement, the absolute branching ratios cannot be extracted. Instead we report relative $p_0/\alpha_0$ branching ratios in Table \ref{tab:RelativeBranchingRatios}. \textcolor{black}{Since we report relative branching ratios, the origin and nature of the background in the inclusive spectrum does not influence our final result.} The yields for these states are assumed to follow a log-normal distribution: the ratio of the branching ratios must then also be log-normally distributed and the low, median and upper values given in Table \ref{tab:RelativeBranchingRatios} are given assuming this.

For many states, no $p_0$ decay is observed above background. For these cases, we estimate what the smallest yield would have been for the state to have been observed, and this yield is used to compute the upper limit for the ratio of the branching ratios. The probability distribution for an observable peak above the background is calculated using the Feldman-Cousins method \cite{PhysRevD.57.3873} using the implementation in the ROOT data-analysis framework \cite{ROOT}. The probability distribution function for the ratio of the $p_0$ and $\alpha_0$ decays is then calculated by Monte-Carlo sampling of the probability distribution functions for the $p_0$ and $\alpha_0$ decays; the probability distribution function for the $\alpha_0$ decay is assumed to follow a log-normal distribution. The resulting numerically determined 84\% percentiles are reported in Table \ref{tab:RelativeBranchingRatios}.

\begin{table}
    \centering
    \caption{Ratio of the proton to the $\alpha$-particle branching ratio for states above the neutron threshold in $^{19}$F, and for states below the neutron threshold which were also measured in the $^{18}$O$+p$ data of Sellin {\it et al.} See the text for details as to how the ratio is computed. The index corresponds to the state indices of Table \ref{tab:Q3DResults} and the `low', `median' and `high' values are the 16\%, 50\% and 84\% percentiles for the ratio assuming that the data follow a log-normal distribution. For those cases where no `low' and `median' values are given, the ratio is only an upper limit given to the 84\% percentile. The final column gives the ratio of the widths determined from the $^{18}$O$+p$ data of Sellin {\it et al.} \cite{SELLIN1969461} though no uncertainties are available for these ratios.}
    \label{tab:RelativeBranchingRatios}
    \begin{ruledtabular}
    \begin{tabular}{c c c c c c c c}
       Index  & \makecell{$E_x$\\~[MeV]} & \makecell{Lower \\ $B_{p_0}/B_{\alpha_0}$} & \makecell{Median\\ $B_{p_0}/B_{\alpha_0}$} & \makecell{Upper\\ $B_{p_0}/B_{\alpha_0}$} & \makecell{$B_{p_0}/B_{\alpha_0}$\\Ref. \cite{SELLIN1969461}} \\
       \hline
         4 & $10.231$ & $1.23$ & $1.63$ & $2.14$ & $1.65$ \\
         5 & $10.256$ & $1.17$ & $1.31$ & $1.45$ & $0.85$ \\
         6 & $10.308$ & $1.47$ & $1.80$ & $2.19$ & $1.16$ \\
        10 & $10.420$ &  & & $0.01$ & & \\
        11 & $10.452$ &  & & $0.07$ & & \\
        12 & $10.486$ &  & & $0.07$ & & \\
        13 & $10.500$ & $1.77$  & $3.42$ & $6.62$ & $2.4$ \\
        14 & $10.515$ &  & & $0.02$ & & \\
        15 & $10.547$ &  & & $0.03$ & & \\
        16 & $10.553$ &  & & $0.02$ & & \\
        17 & $10.570$ &  & & $0.06$ & & \\
        18 & $10.579$ &  & & $0.03$ & & \\
        19 & $10.596$ &  & & $0.07$ & & \\
        20 & $10.616$ & $0.21$  & $0.25$ & $0.28$ & $0.64$ \\
        21 & $10.676$ &  & & $0.07$ & & \\
    \end{tabular}
    \end{ruledtabular}
\end{table}

Table \ref{tab:RelativeBranchingRatios} also includes the ratio  of the $p_0$ and $\alpha_0$ partial widths determined from the $^{18}$O$+p$ data of Sellin {\it et al.} \cite{SELLIN1969461}. There is moderate agreement between the present results and those of Sellin {\it et al.} \cite{SELLIN1969461} though direct comparison is hindered by the lack of uncertainties reported in those results.

\subsection{Discussion}

Due to the considerable uncertainty in the absolute branching ratios resulting from the high background in the singles spectrum from the Orsay Split-Pole, it is not possible to calculate the $^{18}$F($n,\alpha$)$^{15}$N and $^{18}$F($n,p$)$^{18}$O reaction rates with reasonable uncertainties. The neutron widths or branching ratios are not available, and while a Monte-Carlo approach of reasonable values for the widths is possible, the resulting uncertainty in the reaction rates is extremely high. Instead, we limit the discussion to qualitatively considering the relative strengths of the $\alpha$-particle and proton decay modes from the states above the neutron threshold in $^{19}$F.

It is clear from Figure \ref{fig:OrsaySpectra} that the $\alpha$-particle decay branch is significantly stronger than the proton decay branch. In fact, above the neutron threshold, the only states observed to have proton decay branches are at $E_x = 10.500$ and $10.616$ MeV. This result is in partial agreement with the $^{18}$O$+p$ data of Carlson who observed two states strongly in the $^{18}$O($p,p$)$^{18}$O reaction (indices 21 and 24 in that work) at proton bombarding energies corresponding to $E_x = 10.500$ and $E_x = 10.616$ MeV.

\textcolor{black}{The dominance of $\alpha$-particle emission supports an increased $^{15}$N production in the models of core-collapse supernovae by Bojazi and Meyer \cite{PhysRevC.89.025807}, and the cause of the enriched hotspots observed in presolar grains \cite{2041-8205-754-1-L8}. Remaining $^{18}$F will decay into $^{18}$O following the shockwave. However, firm conclusions as to the production of $^{15}$N require data on the neutron widths of states in $^{19}$F, since the ratio of the $^{18}$F($n,\alpha$)$^{15}$N and $^{18}$F($n,p$)$^{18}$O reaction rates depends not only on the relative branching of the $^{19}$F but also their probabilities of population, which are given by the neutron widths.}

\section{Conclusions}

The $^{19}$F($p,p^\prime$)$^{19}$F reaction has been studied using the Q3D magnetic spectrograph at Munich, and the Orsay SplitPole. Excited states above the neutron threshold in $^{19}$F were observed, along with charged-particle decays from those states. The $\alpha$-particle decay of the states above the neutron threshold is generally stronger than the proton decay. However, without any information on the neutron widths the reaction rates cannot yet be calculated. The observation of $\alpha$-particle decay branches from the observed states in $^{19}$F is consistent with the suggestion of Bojazi and Meyer \cite{PhysRevC.89.025807} that the $^{18}$F($n,\alpha$)$^{15}$N reaction could contribute to the production of $^{15}$N in the helium-burning layer of core-collapse supernovae. This, in turn, can explain the spatially correlated over-abundances of $^{15}$N and $^{18}$O observed in some meteoritic grains in the Orgueil meteorite \cite{2041-8205-754-1-L8}.

Calculations of the $^{18}$F($n,p$)$^{18}$O and $^{18}$F($n,\alpha$)$^{15}$N reaction rates are not presently possible due to the lack of information about the neutron widths. Time-reversed measurements, i.e. the $^{15}$N($\alpha,n$)$^{18}$F and $^{18}$O($p,n$)$^{18}$F reactions, would provide the required information and should be the focus of future experimental studies.

\section{Acknowledgements}

It is a pleasure to be able to thank the beam operators at the Maier-Liebnitz Laboratorium, Munich and the ALTO facility of IJCLab for the high-quality beams provided, and the target laboratory staff of the LNS-Catania for their excellent targets. PA thanks Dr Matthew Williams of the University of York and TRIUMF for discussions about Q3D data and the lithium background. This work has been supported by the IN2P3-ASCR LEA NuAG. V.G. would like to thank S\~{a}o Paulo Research Foundation (FAPESP) (Grants No. 16/50315-6 and 14/14432-2). MM thanks the deanship of scientific research, Jazan University for support.

\bibliography{F19_structure}

\end{document}